\documentclass[showpacs,oneside,twocolumn,prl,amsmath,amssymb,superscriptaddress]{revtex4-1}
\usepackage{cases}
\usepackage{amsmath}
\usepackage{amssymb}
\usepackage{amsfonts}
\usepackage{amssymb}
\usepackage{dcolumn}
\usepackage{bm}
\usepackage{bbm}
\usepackage{graphicx}
\usepackage{xcolor}
\usepackage{array}
\usepackage{subfigure}

\usepackage[colorlinks,urlcolor=blue,citecolor=blue]{hyperref}

\usepackage[none]{hyphenat}
\usepackage{CJKutf8}

\begin{document}
\title{Lens-free Optical Detection of Thermal Motion of a Sub-millimeter Sphere Diamagnetically Levitated in High Vacuum }

\author{Fang Xiong  }
\thanks{These authors contributed equally to this paper.}
\affiliation{National Laboratory of Solid State Microstructures and Department of Physics, Nanjing University, Nanjing, 210093, China}

\author{Peiran Yin }
\thanks{These authors contributed equally to this paper.}
\affiliation{National Laboratory of Solid State Microstructures and Department of Physics, Nanjing University, Nanjing, 210093, China}
\author{Tong Wu}
\affiliation{National Laboratory of Solid State Microstructures and Department of Physics, Nanjing University, Nanjing, 210093, China}

\author{Han Xie}
\affiliation{National Laboratory of Solid State Microstructures and Department of Physics, Nanjing University, Nanjing, 210093, China}
\author{Rui Li}
\affiliation{CAS Key Laboratory of Microscale Magnetic Resonance and Department of Modern Physics, University of Science and Technology of China, Hefei 230026, China}
\affiliation{Synergetic Innovation Center of Quantum Information and Quantum Physics, University of Science and Technology of China, Hefei 230026, China}
\affiliation{Hefei National Laboratory for Physical Sciences at the Microscale, University of Science and Technology of China, Hefei 230026, China}
\author{Yingchun Leng}
\affiliation{National Laboratory of Solid State Microstructures and Department of Physics, Nanjing University, Nanjing, 210093, China}
\author{Yanan Li}
\affiliation{National Laboratory of Solid State Microstructures and Department of Physics, Nanjing University, Nanjing, 210093, China}
\author{Changkui Duan}
\affiliation{CAS Key Laboratory of Microscale Magnetic Resonance and Department of Modern Physics, University of Science and Technology of China, Hefei 230026, China}
\affiliation{Synergetic Innovation Center of Quantum Information and Quantum Physics, University of Science and Technology of China, Hefei 230026, China}
\affiliation{Hefei National Laboratory for Physical Sciences at the Microscale, University of Science and Technology of China, Hefei 230026, China}
\author{Xi Kong}
\thanks{Corresponding author: \href{mailto:kongxi@nju.edu.cn}{kongxi@nju.edu.cn}}
\affiliation{National Laboratory of Solid State Microstructures and Department of Physics, Nanjing University, Nanjing, 210093, China}
\author{Pu Huang}
\thanks{Corresponding author: \href{mailto:hp@nju.edu.cn}{hp@nju.edu.cn}}
\affiliation{National Laboratory of Solid State Microstructures and Department of Physics, Nanjing University, Nanjing, 210093, China}
\author{Jiangfeng Du}
\thanks{Corresponding author: \href{mailto:djf@ustc.edu.cn}{djf@ustc.edu.cn}}
\affiliation{CAS Key Laboratory of Microscale Magnetic Resonance and Department of Modern Physics, University of Science and Technology of China, Hefei 230026, China}
\affiliation{Synergetic Innovation Center of Quantum Information and Quantum Physics, University of Science and Technology of China, Hefei 230026, China}
\affiliation{Hefei National Laboratory for Physical Sciences at the Microscale, University of Science and Technology of China, Hefei 230026, China}

\begin{abstract}
Levitated oscillators with millimeter or sub-millimeter size are particularly attractive due to their potential role in studying various fundamental problems and practical applications. One of the crucial issues towards these goals is to achieve efficient measurements of oscillator motion, while this remains a challenge. Here we theoretically propose a lens-free optical detection scheme, which can be used to detect the motion of a millimeter or sub-millimeter levitated oscillator with a measurement efficiency close to the standard quantum limit with a modest optical power. We demonstrate experimentally this scheme on a 0.5 mm diameter micro-sphere that is diamagnetically levitated under high vacuum and room temperature, and the thermal motion is detected with high precision. Based on this system, an estimated acceleration sensitivity of $9.7 \times 10^{-10}\rm g/\sqrt{Hz}$ is achieved, which is more than one order improvement over the best value reported by the levitated mechanical system. Due to the stability of the system, the minimum resolved acceleration  of $3.5\times 10^{-12}\rm g$ is reached with measurement times of $10^5$ s. This result is expected to have potential applications in the study of exotic interactions in the millimeter or sub-millimeter range and the realization of compact gravimeter and accelerometer.
\end{abstract}
\maketitle
\textit{Introduction.} \textbf{---}
Levitated mechanical oscillator in vacuum is supposed to achieve ultra low mechanical noise  due to its isolation from environment \cite{Chang_PNAS_2010}. It has been proposed for studying various fundamental physics such as test of the wave function collapse models \cite{Diosi_PRL_2015,Li_PRA_2016, Vinante_PRA_2019,Zheng_PRR_2020,Pontin_PRR_2020,Vinante_PRR_2020} and investigation of new physics \cite{Geraci_PRL_2010,Arvanitaki_PRL_2013,Moore_PRL_2014,Dark_Rider_PRL_2016,Kawasaki_PRD_2019, Dark_Matter_Monteiro_PRL_2020, Carney_PRD_2020}. For metrology study, it has also been applied in the force detection \cite{force_Gieseler_NP_2013,force_Ranjit_PRA_2016,force_Hempston_APL_2017}, the inertia sensing \cite{superconducting_gravimeter_Goodkind_1999,Moody_2002,accelerometer_superconductor_Chirs_2019,accelerometer_optical_Monteiro_PRA_2017,coat_goldmirror_diamagnetic2020,63um_magnetic_gravity_high_sensitivity_2021}, the thermometry \cite{thermometry_Millen_NN_2014,thermometry_Hebestreit_PRA_2018}, and the magnetometry \cite{Magnetometer_Jackson_PRL_2016,Magnetometer_Band_PRL_2018}. According to different applications, the sizes of the levitated mechanical oscillators range from  26 nm  using room temperature optical trap \cite{optical_Vovrosh_JOS_2017} to centimeter level using a superconducting levitation with mass greater than 1 kg \cite{Moody_2002}.

Acceleration sensitivity is a crucial issue in many researches. The  centimeter level superconducting levitated oscillator has been demonstrated to have an ultra-high acceleration sensitivity of $\sim10^{-10}\rm g /\sqrt{Hz}$  \cite{superconducting_gravimeter_Goodkind_1999}, and sub-millimeter superconducting levitated microsphere has been reported with a potential acceleration sensitivity order of $ 1.2 \times 10^{-10}\rm g/\sqrt{Hz}$, assuming that the thermal noise limit is reached \cite{accelerometer_superconductor_Chirs_2019}. Smaller oscillator with the diameter less than tens of micrometers have been studied in systems based on optical levitation \cite{accelerometer_optical_Monteiro_PRA_2017}, diamagnetical levitation \cite{Diamagnetic_10um_IEEE2008,magneto_gravitational_20um_RSI2018,63um_magnetic_gravity_high_sensitivity_2021}, and  superconducting levitation \cite{10umsupercounduct_Cirio_PRL_2012,12umSuperconductor_Wang_PRApp_2019,15umsuperconduct_Gieseler_PRL_2020}. The best reported levitated oscillator acceleration sensitivity of  $3.6\times\rm 10^{-8}\rm g/\sqrt{Hz}$ has been achieved at room temperature  \cite{63um_magnetic_gravity_high_sensitivity_2021} with magnetic gravitational system.

The mechanical oscillator based acceleration sensing is characterized by the power density of the acceleration noise $S_{\rm aa}^{\rm tot} = S_{\rm aa}^{\rm th} + S_{\rm aa}^{\rm mea}$. $S_{\rm aa}^{\rm th} = 4 \gamma k_{B}T/m$ is the thermal Brownian noise, where $m$ is the system mass, $\gamma/2\pi$ is the dissipation rate of the oscillator and $T$ is the environment temperature. The thermal Brownian noise set a low limit of the achievable sensitivity when in the thermal noise limit. $S_{\rm aa}^{\rm mea}$ is the measurement noise, which is important in practical realization, and in principle is limited by quantum mechanics \cite{RevModPhys_quantum_noise_2010}. Recent experiments using optically levitated nano-oscillator has been shown to be close to oscillator's quantum zero-point motion \cite{optical_Tebbenjohanns_PRL_2020,optical_Deli_science_2020}.

Millimeter or sub-millimeter mechanical oscillators with milligram masses have recently attracted theoretical interest \cite{Quantum_2020}. One of the main advantages of these systems is that the levitated objects can be manufactured as optical components. Combined with the promising methods such as optical radiation pressure suspension \cite{Scattering_Free_PRL_2013,Optics_Express_2017}, diamagnetic levitation\cite{Diamagnetic_Nakashima_PLA_2020,coat_goldmirror_diamagnetic2020} and superconducting levitation \cite{Superconduct_mirror}, it is possible to make the measurement efficiency close to the standard quantum limit (SQL). However, it is still challenging to experimentally achieve a high measurement efficiency of millimeter or sub-millimeter levitation oscillators under high vacuum conditions \cite{Superconduct_mirror}. Especially, the observation of the thermal Brownian motion of millimeter or sub-millimeter levitation oscillators is still elusive.

In this  experiment, we propose a lens-free optical detection scheme that can detect the motion of the diamagnetically levitated transparent micro-sphere which can be made of quartz, diamond and most organic compounds like Polymethyl methacrylate (PMMA). It can be used as  a mechanical oscillator or as an optical functional element for motion measurement. Theoretically, this method can achieve the detection efficiency close to the SQL of sub-millimmeter micro-sphere through moderate laser power. Then we demonstrate the scheme by using diamagnetically levitated sphere made of PMMA of diameter 0.5 mm at room temperature and the thermal Brownian motion is detected.  Under high vacuum, we experimentally measured system's total acceleration noise, which corresponds to the estimated sensitivity of $ (9.7 \pm 1.1) \times 10^{-10}\rm\/ g/\sqrt{\rm Hz}$, more than an order of magnitude improvement over the best reported levitation oscillator at room temperature. The  estimated minimum resolved acceleration of $(3.5 \pm 1.4)\times 10^{-12}$ g is achieved through the measurement time about $10^5$ s, comparable to the best reported value based on lab-scale system of cold atom interferometry \cite{Luo_cold-atom_2013,Kasevich_cold-atom_2020}. Further improvements and potential applications of the system will be discussed.

\begin{figure}
\includegraphics[width=8.6cm]{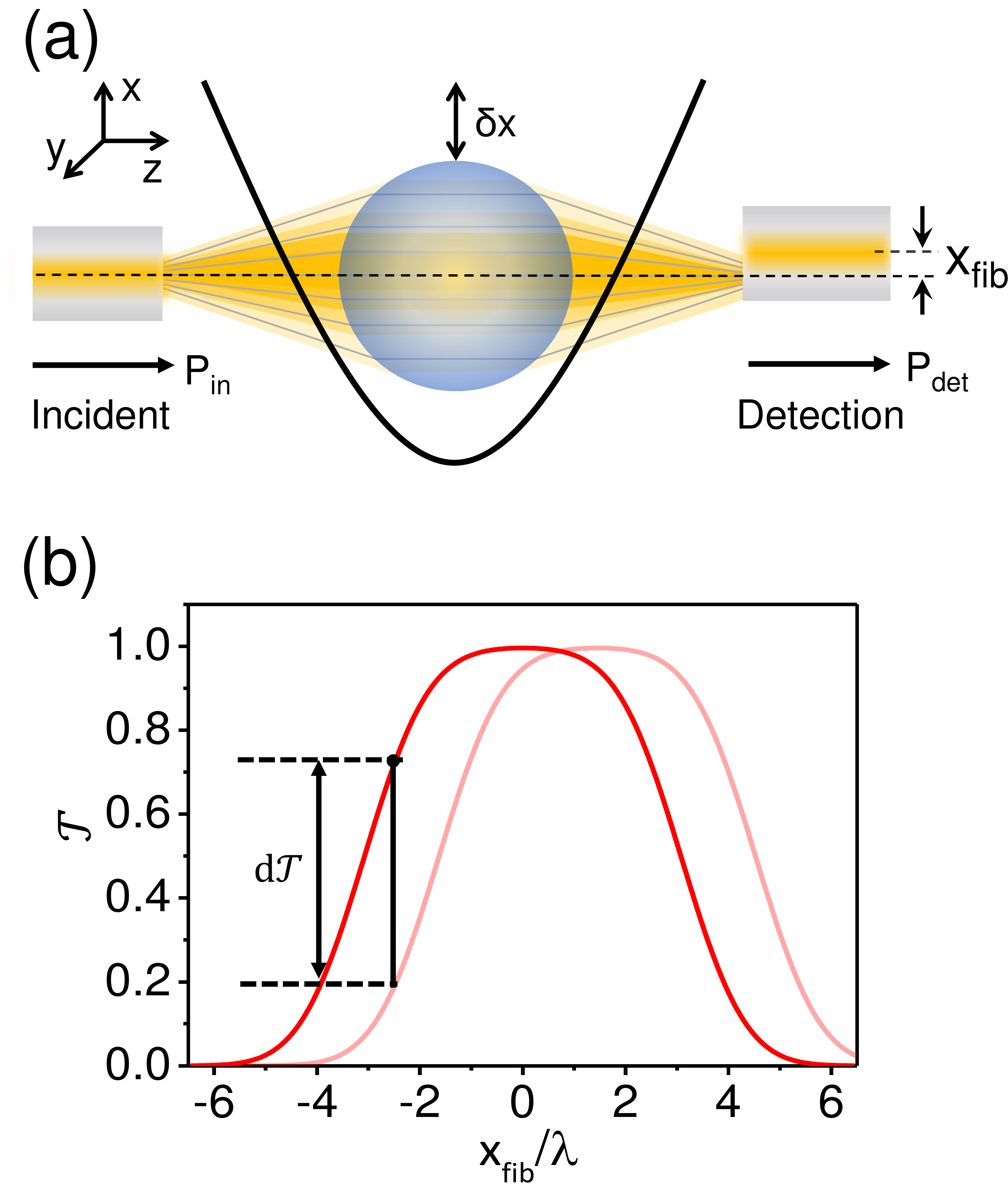}
\caption{ Schematic of our method. (a) Lens-free optical measurement system. The 1550 nm laser with a power of $ P_{\rm in} $ illuminates the levitated micro-sphere via the left incident fiber. The micro-sphere works as a spherical lens to refocus the light into the detection fiber on the right. The optical fiber is positioned $ x_{\rm fib} $ relative to the center in the $x$ axis direction, and the power detected by the detection optical fiber is $ P _ {\rm out} $. (b) The calculated transmission coefficient  $\mathcal{T} = P_{\rm out} / P_{\rm in} $ as a function of the shift of detecting fiber position $ x_{\rm fib} $. The red line is the curve where the micro-sphere is in the central equilibrium position, and the light red curve corresponds to the response curve a small displacement $ \delta x $ along the $x$ axis. The black line represents the response of signal $ \delta \mathcal{T} $ of the measured signal to the displacement $ \delta x $.
}
\label{fig1}
\end{figure}

\textit{Theoretical description of the method } \textbf{---}
The key idea of our detection method is based on the optical property of the  micro-sphere as a spherical lens, shown in Fig.~\ref{fig1}(a), where a levitated transparent micro-sphere is placed between two fibers. The light comes from the end of a incident fiber with the power $P_{\rm in}$, and the micro-sphere is used as a spherical lens to refocus the light to the end of another detection fiber. Thus the collected optical power $P_{\rm det}$ depends on the position of the micro-sphere. The transmission coefficient $\mathcal{T} = P_{\rm det}/P_{\rm in}$ is sensitive to the displacement of the micro-sphere, thus providing a new method to measure the motion of the oscillator.  Fig.~\ref{fig1}(b) shows the response of transmission coefficient $\mathcal{T}$ to the detection fiber's position $x_{\rm fib}$  under different micro-sphere displacements $\delta x$ along $x$ axis. By adjusting the detection fiber's position $x_{\rm fib}$ , an optimal sensitivity can be achieved \cite{SM}.

Our scheme provides a potential capability to reach a high measurement efficiency of the micro-sphere motion \cite{SM}.
The motion equation of a classic oscillator subject to noises is,
\begin{align}
\label{EOM} m\ddot{x}+m\gamma\dot{x}+ m \omega_{\rm{0}}^2 x=
f_{\rm{th}}(t)+ f_{\rm{ba}}(t),
\end{align}
where $m$ is the oscillator mass, $\gamma/2\pi$ is the mechanical dissipation rate, $\omega_{\rm{0}}/2\pi$ is the resonant frequency, $f_{\rm{th}}(t)$ is the  thermal Brownian noise and $f_{\rm{ba}}(t)$ is the back action fluctuation, which comes from the photon shot noise of incident laser on the oscillator. Thus the detected power spectral density (PSD) of displacement of oscillator due to Brownian noise and back action fluctuation is $S_{\rm xx}^{\rm th}(\omega) = 4\gamma k_{\rm B}T |\chi(\omega)|^2 / m$ and  $S_{\rm xx}^{\rm ba}(\omega) = |\chi(\omega)|^2 S_{\rm FF}^{\rm ba}(\omega)/m^2$ correspondingly, where $\left|\chi\left(\omega\right)\right|^2=1/[\left(\omega_{\rm{0}}^2-\omega^2\right)^2+\gamma^2\omega^2]$ is the mechanical susceptibility. The measurement imprecision $S_{\rm xx}^{\rm imp}$ and backaction satisfied relation $ S_{\rm xx}^{\rm imp}S_{\rm FF}^{\rm ba}  = (1/\eta)\hbar^2$, here $\eta \leq 1$ is measurement efficiency, and detection reached standard quantum limit (SQL) for $\eta = 1$. Considering the measurement imprecision, the PSD of the displacement of the oscillator is $S_{\rm xx}^{\rm tot}(\omega) = S_{\rm xx}^{\rm th}(\omega) +S_{\rm xx}^{\rm imp}+S_{\rm xx}^{\rm ba}(\omega)$. The measurement related noises, the back-action noise $S_{\rm xx}^{\rm ba}$ and the imprecision noise $S_{\rm xx}^{\rm imp}$, can be further optimized. The measurement related noise $S_{\rm aa}^{\rm mea}$ of the acceleration is \cite{SM},
\begin{align}
\label{acceleration noise}S_{\rm aa}^{\rm mea} = \frac{S_{\rm xx}^{\rm imp}}{|\chi(\omega)|^2} + \frac{S_{\rm FF}^{\rm ba}}{m^2} .
\end{align}

\begin{figure}
\includegraphics[width=8.6cm]{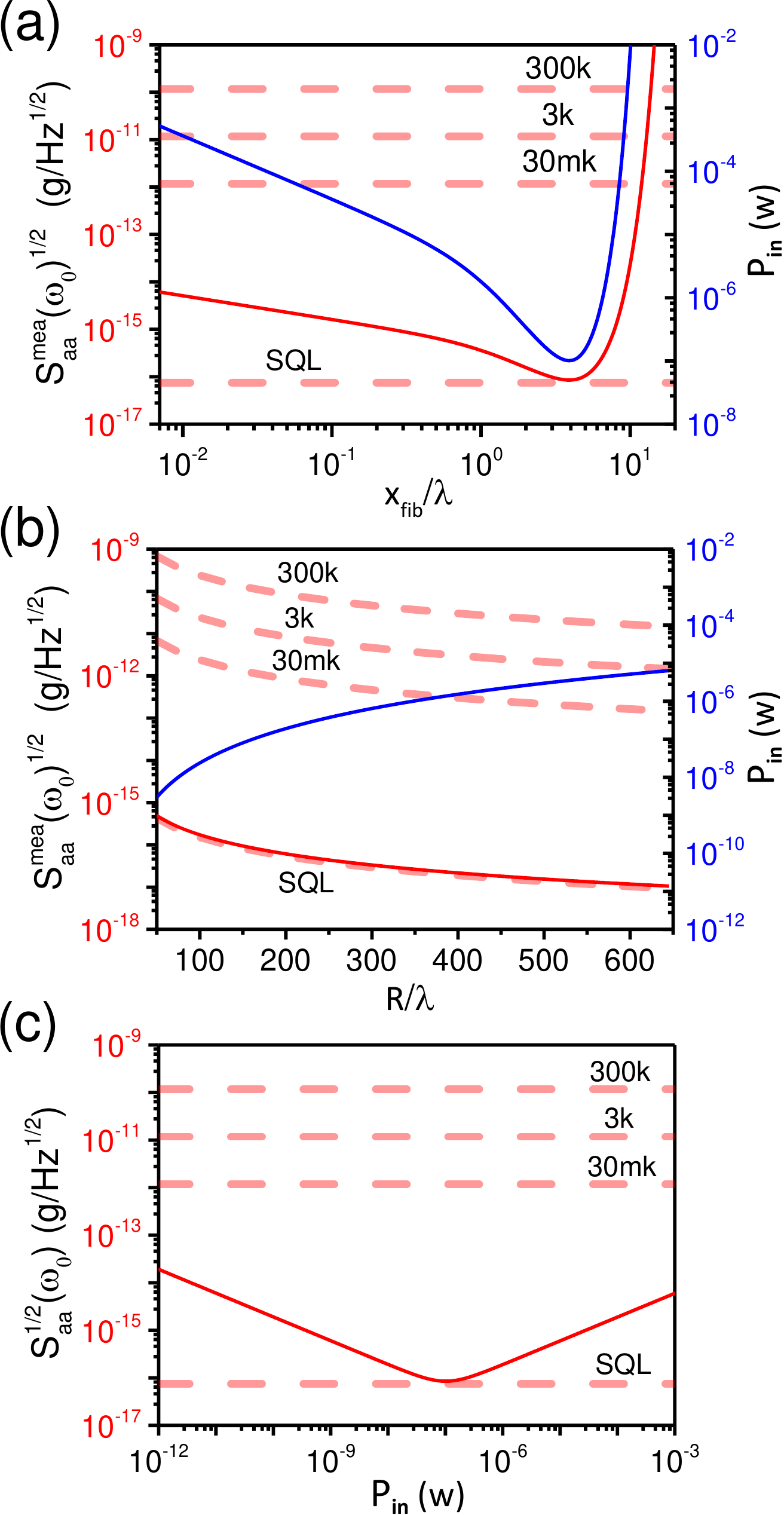}
\caption{
Estimated measurement noise of acceleration. (a) The red curve (left axis) is the minimum measurement noise of acceleration $\sqrt{S_{\rm aa}^{\rm mea}}$ versus the shifted detection fiber position $x_{\rm fib}$. We assume that the refractive index of the micro-sphere $ n = 1.4 $, the diameter $ 2R = 0.5~$mm, the acceleration noise is estimated using the mode that oscillates along the $x$ axis, the resonant frequency is 10~Hz, and the mechanical dissipation is $ 10^{-6}~$Hz. The red dashed line is the thermal noise of acceleration $\sqrt{S_{\rm aa}^{\rm mea}}$ and standard quantum limits (SQL) as labeled. The blue curve (right aixs) is the corresponding optimized laser power $ P_{\rm in}^{\rm opt} $. (b) Same as (a) but versus the size of microsphere (radius $R$ in unit of wavelength) with a fixed $x_{\rm fib}=5.8 \mu m$. (c) Same as (a) but versus the incident laser power, and with a fixed $x_{\rm fib}=5.8 \mu m$.}
\label{fig2}
\end{figure}

As an example, we consider a levitated oscillator with a micro-sphere diameter of $0.5$ mm, relative index of refraction $n = 1.4$, resonant frequency $\omega_0/2\pi = 10$ Hz and mechanical dissipation $\gamma/2\pi = 10^{-6}$ Hz.  As shown in Fig.~\ref{fig2} (a), the minimum $S_{\rm aa}^{\rm mea}$ can be obtained by optimizing the detection fiber position $x_{\rm fib}$ and the incident power $P_{\rm in}$.  Fig.~\ref{fig2} (b) shows the minimum measurement noise of acceleration as a function of different micro-sphere radiuses as well as optimized laser powers.

In practical applications, the low frequency oscillator is not necessarily close to the quantum limit, because the system remains far from the quantum regime even at low temperature. In particular,  since the micro-sphere absorbs light and heats up under high vacuum conditions, the laser power may become a crucial issue. The additional noise \cite{accelerometer_optical_Monteiro_PRA_2017} brought by the micro-spheres is calculated and plotted in Fig.~\ref {fig2}(c) under different laser power at the optimal detection position $ x _{\rm fib} $. By measuring the noise, we find that even at ultra-low temperature and the laser power range from $ 10 ^ {-12} $ W to $ 10 ^ {-3} $ W, the measurement noise is still much lower than the thermal noise.


\begin{figure}
\includegraphics[width=8.6cm]{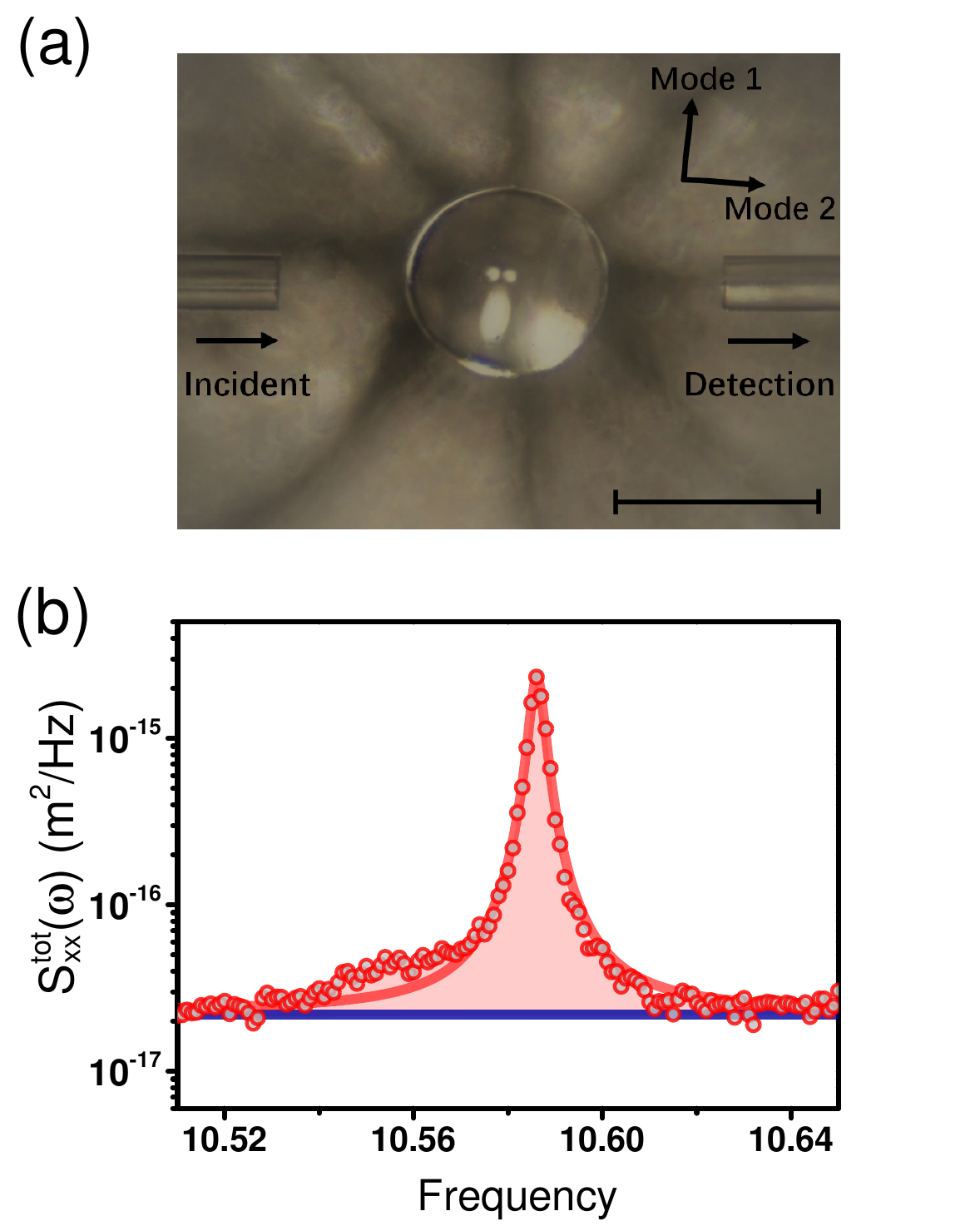}
\caption{
Experimental measurement of thermal movement of sub-millimeter micro-sphere. (a) The optical image of a micro-sphere with $0.5$ mm diameter levitated  in a magnetic gravitational trap. A 1550 nm laser light transmits  via the left fiber to the right one passing the sub-millimeter micro-sphere. The fibers are fixed on magnetic trap by UV glues. The vibration mode 1 and mode 2 on the horizontal plane are indicated by arrows on the inset figure, and mode 1 is used in the current experiment. The angle between Mode 1 and the optical axis is about $85 \pm 5$ degrees.(b) The measured  power spectral density of motion $S_{\rm xx}^{\rm tot} (\omega)$ of oscillation mode 1 under the pressure of $4 \times 10^{-3}$ mbar (red dots). The data is fitted to the Lorentz curve (red line). The blue line shows the fitted imprecision noise power spectral density $S_{\rm xx}^{\rm imp} $. The light red area represents the contribution of effective thermal Brownian motion at an ambient temperature of 298K.  }
\label{fig3}
\end{figure}

\textit{Experimental demonstration} \textbf{---}
We test our scheme using a diamagnetically levitated oscillator at room temperature. A 0.5 mm diameter  methyl methacrylate micro-sphere is suspended in a magnetic trap \cite{Zheng_PRR_2020}. A group of NdFeB magnets of octagonal bilayer geometry with gravity generated stable diamagnetic levitation potential. Especially, two narrow grooves is fabricated on the magnets  so fibers can go to the trapping area as shown in Fig.~\ref{fig3}(a). After loading the micro-sphere into the trap, a 1550nm laser with the power order of $\mu$W is applied to the incident fiber while the photon detector is connected to the detection fiber. A pair of 3-axis piezoelectric positioners are used to adjust the position of the incident and detection optical fiber. After reaching the optimal position, the optical fiber is fixed to the magnetic trap by UV glues, and the piezoelectric positioner is removed to avoid unnecessary vibration. The device is placed in a vacuum chamber with a spring-mass suspension based vibration isolator. The whole system is placed in a home-build thermostatic container. During the experiment, the performance fluctuation of the container can be controlled below 50 mK.

According to the design of the magneto-gravitational trap, the micro-sphere oscillator has three orthogonal oscillation modes, one mode oscillates along the direction of gravitational field (mode $g$), and the other two modes oscillate in the horizontal plane (denoted as mode 1 and mode 2). Oscillation mode 1 is vertical to the optical axis, and the mode 2 is along the axial direction. In experiment, as shown in Fig.~\ref{fig3}(a), the imperfection of the magneto-gravitational trap and the  micro-sphere make the mode $g$ , mode 1 not perfectly vertical to and  mode 2 not perfectly axial to the optical axis. Without loss of generality, we study the vertical mode (mode 1) whose oscillation direction is close to the vertical direction of the optical axis as shown in Fig.~\ref{fig3}(a), and the corresponding resonance frequency is 10.58~Hz.

In the experiment, the pressure of the vacuum chamber is $4 \times 10^{-3}$ mbar. The motion of the micro-sphere oscillator is measured by a photon detector, and the voltage signal $V(t)$ is then amplified and recorded. The measurement time is 500 s, which is much longer than the relaxation time of the oscillator. The corresponding power spectral density is $S_{V}(\omega) = \xi^2 S_{\rm xx}^{\rm tot}(\omega)$, where $\xi$ is the displacement-voltage conversion coefficient. Considering the detection bandwidth $b$, the detected signal is  $\int_{\omega_0-b/2}^{\omega_0+b/2} S_{V}(\omega)d\omega$. The effective temperature $T_{\rm eff}$ of the oscillator is defined using the measured power spectral density of motion as $ T_{\rm eff} = (m \omega_{0}^2 / k_{\rm B})   \int_{\omega_0-b/2}^{\omega_0+b/2} \left(S_{\rm FF}^{\rm th}+S_{\rm FF}^{\rm ba}\right)|\chi(\omega)|^2/m^2 d\omega$ \cite{SM}. Then we obtained a simpler expression  as $\int_{\omega_0-b/2}^{\omega_0+b/2} S_{V} (\omega)d\omega = \xi^2 ( k_{\rm B}T_{\rm eff}/m\omega_0^2 + b S_{\rm xx}^{\rm imp}  )$. Under  current pressure ($10^{-3}$ mbar), the oscillator is fully thermalized with the environment, so we take the effective temperature $T_{\rm eff} = T_{\rm en} =298$ K  measured by a thermometer fixed on the vacuum chamber. So by fitting the measured power spectral density $S_{V}(\omega) $ to Lorentz curve and considering the frequency independent baseline $\xi^2 b S_{\rm xx}^{\rm imp} $, conversion coefficient $\xi $ and imprecision noise $S_{\rm xx}^{\rm imp}$ are obtained, and Fig.~\ref{fig3}(b) shows the power spectral density of displacement $S_{\rm xx}^{\rm tot}(\omega) = S_{V} (\omega)/\xi^2  $.

Then we pumped the chamber to a lower pressure of $1.2 \times 10^{-5}$ mbar to study the acceleration sensitivity of the oscillator. We first measure the effective temperature $T_{\rm eff}$ by measuring the power spectral density of the displacement. The mechanical dissipation coefficient obtained from free ringing at lower pressure is as low as $\gamma/2 \pi = 71\rm$ $\mu$Hz, and the corresponding correlation time is $\tau \approx 2200$ s. Because of the statistical error of the effective temperature $\Delta T_{\rm eff} \propto \sqrt{1/t_{\rm mea} \gamma}$ (see \cite{SM}), our measurement time in the experiment is $t_{\rm mea} = 7\times10^5$ s\ $\gg \tau $ (about 190 hours). We have obtained the effective temperature $T_{\rm eff} = 289 \pm 67$ K which is consistent with the environment temperature. The results show that the oscillator is still in the thermal equilibrium, and external noises, such as laser heating and vibration, can be neglected \cite{SM}. Therefore, the pressure is not further reduced, because the mechanical dissipation $\gamma/2\pi$ is saturated at the current pressure, while such dissipation is beyond the scope of current research. We estimate the power spectral density of acceleration noise as
\begin{align}
\label{acceleration} S_{\rm aa}^{\rm tot}(\omega) = \frac{4 \gamma k_{\rm B} T_{\rm eff} }{m} + \frac{S_{\rm xx}^{\rm imp}}{|\chi(\omega)|^2}.
\end{align}
Based on the current experiment, the total acceleration noise at the resonant frequency is obtained as $\sqrt{S_{\rm aa}^{\rm tot}(\omega_0)} = (9.7 \pm 1.1) \times 10^{-10} \rm g/\sqrt{Hz} $. The related results are summarized in Table \ref{tab:table2}.

\begin{figure}
\includegraphics[width=9.8cm]{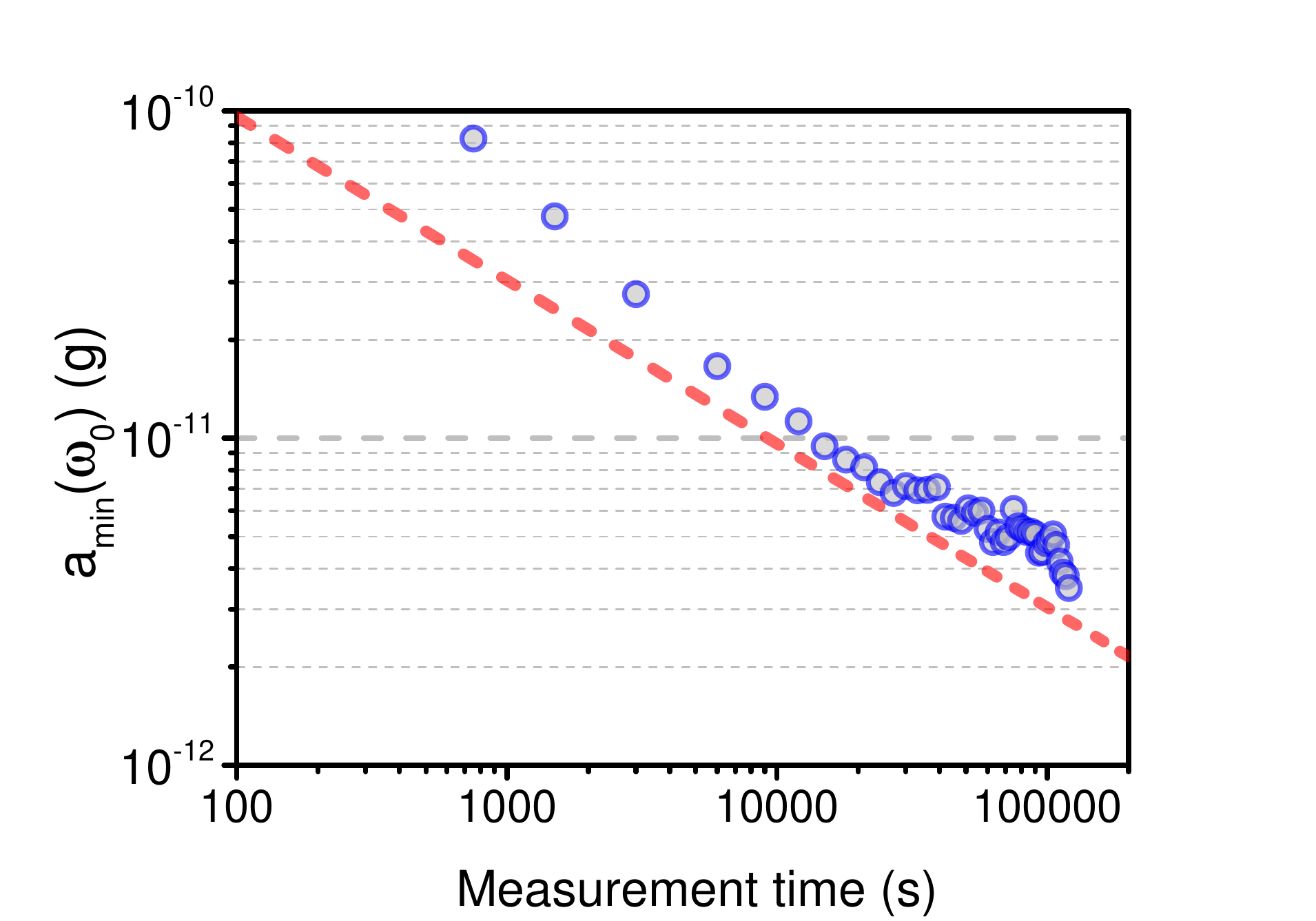}
\caption{
The minimum resolvable acceleration. The blue dot is the measured acceleration noise as a function of measurement time at pressure $1.2 \times 10^{-5}$ mbar. To build up the statistics, the total experiment time is about $7\times 10^5$ s (about 190 hours) has been taken. The red curve is the theoretical curve assuming the temperature fluctuations are  eliminated.  }
\label{fig4}
\end{figure}

Finally, we estimate the minimum resolvable acceleration $a_{\rm min}(\omega_0)$ at the resonant frequency. Considering the oscillating acceleration signal at the resonant frequency, Fig.~\ref{fig4} shows the corresponding $a_{\rm min}(\omega_0)$ as a function of the total measurement time, where $a_{\rm min}(\omega_0) = (3.5 \pm 1.4)\times10^{-12}$ g has reached its minimum value when the measuring time is $10^5$ s. The measured $a_{\rm min}(\omega_0)$ is slightly higher than the predicted value $a_{\rm min}(\omega_0) = \sqrt{S_{\rm aa}^{\rm tot}(\omega_0)/t}$. The possible reason for this difference is that the temperature fluctuation limits the frequency fluctuation $\delta \omega_0/2\pi \approx 770 \mu$Hz, about an order larger than mechanical dissipation $\gamma/2\pi$ (for a detailed description of the data, please refer to SM \cite{SM}). In principle, this can be overcome by better temperature control schemes, such as placing the system in a low temperature environment \cite{Leng_arXiv_2020}.

\begin{table}[h]
	\caption{\label{tab:table2}  Summary of main experiment parameters.
Diameter of micro-sphere is $0.5$ mm  and the
corresponding mass is $\rm 78$  $\mu$g, resonant frequency is $\omega_{0}/2\pi = 10.58$ Hz,  environment temperature is maintained at $298$ K, pressure is  $1.2\times10^{-5}$ mbar.}
	\begin{ruledtabular}
		\begin{tabular}{ccc}
			 Symbol   &  Value&  Unit   \\ \hline
              $\xi  $      &$1.14 \pm 0.16$      &  $10^{10}$ V/m                                  \\ 
			 $\gamma/2\pi$    &$7.1\pm 0.03 $   &  $10^{-5}$Hz  \\ 
             $Q$    &$1.49\pm0.0063$   &  $10^{5}$  \\ 
			 $T_{\rm eff}$    &$289 \pm 67 $   &  K  \\ 
			 $\sqrt{S_{\rm x}^{\rm imp}}$    &$4.70 \pm 0.36 $   &  $10^{-9} \rm m/\sqrt{Hz}$  \\ 
			 $\sqrt{S_{\rm aa}^{\rm imp} (\omega_{0})}$  &  $2.0 \pm 0.2$   &  $10^{-12} \rm g/\sqrt{Hz}$  \\ 
			 $\sqrt{S_{\rm aa}^{\rm tot}(\omega_{0})}$    &$9.7 \pm 1.1 $   &  $10^{-10} \rm g/\sqrt{Hz}$  \\ 
			 $a_{\rm min}(\omega_{0})$\footnote{at measurement time of $ 10^5$s }   &$3.5\pm 1.4 $   &  $10^{-12} \rm g$  \\ 
		\end{tabular}
	\end{ruledtabular}
\end{table}

\textit{Discussion and summary} \textbf{---} We propose a detection scheme for levitated millimeter and sub-millimeter micro-spheres that can theoretically approach the standard quantum limit. Our experiments demonstrate that the use of diamagnetic levitated micro-sphere under high vacuum and the measurement noise is low enough to detect the thermal motion of the micro-sphere. Our method provided alternative way to realized quantum limited displacement measurement for mechanical oscillator, in comparison with mirror based system like Michelson interferometry \cite{SM}. At the same time, the measurement noise is still greater than the theoretical estimation, and the main causes  include the possible misalignment of the incident and detection fibers, the imperfection of the micro-sphere shape, and the intensity noise of the laser source. These problems are expected to be significantly overcomes by further improvement of technology.

Acceleration sensitivity of the current system is also experimentally characterized and a better sensitivity is reached compared with reported mechanical system at room temperature. The experiment system reported here has further potential applications, such as gravimetric analysis and acceleration measurement. Although our system can only measure the relative acceleration, rather than the absolute acceleration as in the cold-atom system, the lens-free design reduces the size of the device, resulting in a more compact system. Due to the sub-millimeter scale and the ultra-low resolvable acceleration of the system, it can also be used to test some fundamental physical models such as dark energy \cite{Khoury_Chameleon}.

\begin{acknowledgments}

\textit{Acknowledgments} \textbf{---} We thank Zhujing Xu, Zhang-qi Yin for helpful discussions. This work was supported by the National Key R\&D Program of China (Grant No.~2018YFA0306600), the National Natural Science Foundation of China (Grant No.~61635012, No.~11890702, No.~12075115, No.~81788101, No.~11761131011, No.~11722544 and No.~ 12075115),
the CAS (Grant No.~QYZDY-SSW-SLH004, No.~GJJSTD20170001, and No.~Y95461A0U2),  and the Anhui Initiative in Quantum Information Technologies (Grant No.~AHY050000).
\end{acknowledgments}


%

\end{document}